\documentclass[12pt,preprint]{aastex}

\shorttitle{Coronal Seismology and Acoustic Waves}
\shortauthors{Klimchuk, Tanner, and DeMoortel}

\received{2004 June 16}
\begin{document}

\title{Coronal Seismology and the Propagation of Acoustic Waves Along Coronal Loops}

\author{J. A. Klimchuk and S. E. M. Tanner \altaffilmark{1}} 
\affil{ Space Science Division, Naval Research Laboratory, Washington, DC 20375}
\email{klimchuk@nrl.navy.mil} 

\author{I. De Moortel} 
\affil{School of Mathematics and Statistics, 
University of St. Andrews, St. Andrews, KY16 9SS, UK }

\altaffiltext{1}{also Institute for Computational Sciences and Informatics, 
George Mason University, Fairfax, VA, 22030}

\begin{abstract}
We use a combination of analytical theory, numerical simulation, and data analysis to study the propagation of acoustic waves along coronal loops.  We show that the intensity perturbation of a wave depends on a number of factors, including dissipation of the wave energy, pressure and temperature gradients in the loop atmosphere, work action between the wave and a flow, and the sensitivity properties of the observing instrument.  In particular, the scale length of the intensity perturbation varies directly with the dissipation scale length (i.e., damping length) and the scale lengths of pressure, temperature, and velocity.  We simulate wave propagation in three different equilibrium loop models and find that dissipation and pressure and temperature stratification are the most important effects in the low corona where the waves are most easily detected.  Velocity effects are small, and cross-sectional area variations play no direct role for lines-of-sight that are normal to the loop axis.  The intensity perturbation scale lengths in our simulations agree very well with the scale lengths we measure in a sample of loops observed by {\it TRACE}.  The median observed value is $4.35\times10^9$ cm.  In some cases the intensity perturbation increases with height, which is likely an indication of a temperature inversion in the loop (i.e., temperature that decreases with height).  Our most important conclusion is that thermal conduction, the primary damping mechanism, is accurately described by classical transport theory.  There is no need to invoke anomalous processes to explain the observations.

\end{abstract}


\notetoeditor{Would you please stack the double panels in Figures 6 and 8 vertically instead of horizontally?}

\section{Introduction} \label{sec:intro}

Recent observations of oscillations in coronal loops have inspired a great deal of interest in the emerging field of ``coronal seismology." See, for example, the proceedings of the {\it SOHO}-13 Workshop \citep{ebf04} and the NATO Workshop on Turbulence, Waves and Instabilities \citep{epra03}.  The aim of coronal seismology is to uncover the properties of the corona by studying their influence on waves.  These properties include the temperature and density structure, the magnetic field structure, and the magnitudes of important transport coefficients.

Much of the work so far has been concerned with standing waves---either transverse standing waves that are revealed as oscillatory displacements of coronal loops and are believed to be excited by large flares (e.g., \citealt{sat02}) or longitudinal standing waves that are revealed as Doppler shifts of very hot ($> 6$ MK) spectral lines and believed to be excited by small flares (e.g., \citealt{wetal02}).  Both types of standing waves are observed to decay on time scales of order 1000 s.  \cite{netal99} and \cite{oa02} concluded that such rapid damping of transverse oscillations implies a dissipation by shear viscosity or electrical resistivity that is many orders of magnitude larger than classical transport theory predicts (e.g., \citealt{b65}).  In contrast, \cite{ow02} and \cite{mes04} concluded that classical thermal conduction is adequate to damp the longitudinal waves.

The work we present here examines the possibility of using {\it propagating} longitudinal waves for coronal seismology.  Specifically, we consider acoustic waves traveling along coronal loops.  It is commonly understood that the amplitude of such waves is decreased by dissipation of the wave energy (damping), but it is not widely appreciated that the amplitude can change for a variety of other reasons.  Furthermore, these additional effects can actually cause the amplitude to increase as the wave propagates.  As a simple illustration, consider an acoustic wave propagating along a static coronal loop in the absence of dissipation.  The energy flux of the wave is then conserved, so
\begin{equation} 
\varepsilon C_s A = {\rm constant} , \label{eq:18} 
\end{equation}
where $C_s = (\alpha P / \rho)^{1/2} \propto T^{1/2}$
is the sound speed (with $\alpha$ ranging from 1 for isothermal waves to 5/3 for adiabatic waves), $A$ is the cross-sectional area of the loop, and 
\begin{equation} 
\varepsilon = \frac{1}{2} \rho \delta v^2    \label{eq:19} 
\end{equation}
is the wave's energy density.  Here, $\delta v$ is the velocity perturbation amplitude, and $P$, $\rho$, and $T$ are the pressure, mass density, and temperature, respectively.  Making use of the ideal gas law, $P = R \rho T$, we obtain
\begin{equation} 
\delta v \propto \frac{T^{1/4}}{\left(PA\right)^{1/2}} , \label{eq:deltav} 
\end{equation}
which shows that the velocity amplitude will vary as the wave propagates whenever $T$, $P$, or $A$ are nonuniform along the loop.  An upwardly propagating wave typically experiences increasing temperatures and decreasing pressures, and both act to increase the velocity amplitude.  This is not a new result.  The direct influence of gravitational stratification on acoustic wave propagation has been studied by \cite{ond99}, \cite{nvbr2000}, \cite{ons00}, and  \cite{dh04}.  Stratification can also have an indirect effect by modifying the dissipation rate \citep{mes04}.

The assumption of no dissipation in the above example is unrealistic.  Linear theory indicates that acoustic waves propagating in the corona will be significantly damped, primarily by thermal conduction and to a lesser extent by compressive viscosity and radiation \citep{pks94a,dh03}.  Unfortunately, the theory has limited validity in the transition region and lower corona, where the temperature and pressure can deviate significantly over the distance of a wavelength.  To study wave damping in this situation requires a full nonlinear treatment (i.e., numerical simulation).  We here consider nonlinear damping by thermal conduction in a non-isothermal loop environment.  See \cite{ons00} for a treatment of nonlinear damping by compressive viscosity in an isothermal plume environment.

Equation \ref{eq:deltav} concerns the velocity perturbation of the wave, yet it is the intensity perturbation that is more typically observed.  Small-amplitude ($4.1 \pm 1.5$ \%) intensity disturbances are commonly seen propagating upward along the lower legs of long coronal loops.  They have been detected by {\it TRACE} \citep{nah99,bc99,diw2000,rvbhpn2001,dhiw2002a} and by the Extreme ultraviolet Imaging Telescope (EIT) on the {\it Solar and Heliospheric Observatory (SOHO)} \citep{bc99}.  Similar disturbances have been observed in polar plumes \citep{oetal97,dg1998}.  Although the disturbances could be transient flows induced by impulsive heating \citep{rpsdg2000}, it is far more likely that they are traveling acoustic waves.  They are periodic over several cycles, with periods of $282 \pm 93$ s, and they have propagation speeds of $122 \pm 43$ km s$^{-1}$ that are comparable to the expected sound speed \citep{dhiw2002a}.  Alfv\'{e}n waves can be ruled out because they would be much faster and because they are incompressible in the linear regime and would not produce the observed intensity perturbations.  

The goal of our study is to evaluate whether these or other acoustic wave observations are useful for coronal seismology.  Can we infer the thermodynamic and geometric structure of loops by measuring the variation of the intensity perturbations?  Can we determine whether thermal conduction is anomalously large in the corona, as appears to be the case for shear viscosity?  The answer has important implications for the microphysics of the Sun and of astrophysical plasmas in general.
  
Our study involves several interrelated parts, and the paper is organized accordingly.  We begin in Section 2 with an analytical treatment of acoustic wave propagation in a generic loop for which the temperature, pressure, and velocity structure is arbitrary.  We derive an expression relating the scale length of the intensity perturbation to the scale lengths of the physical loop parameters.  In Section 3, we apply the results to three specific equilibrium loop models and determine which aspects of the models most influence the intensity perturbation.  We then simulate the propagation of waves in the three loops in Section 4.  By combining the numerical and analytical results, we are able to deduce the classical thermal conduction damping length throughout the loop, including those regions where linear theory breaks down.  In Section 5, we reanalyze the {\it TRACE} observations of \cite{dhiw2002a} and compare them directly with the wave simulations.  We conclude with some final remarks in Section 6.

\section{Analytical Treatment}

As we have indicated, the observed intensity perturbation of an acoustic wave can evolve because of dissipation of the wave energy, nonuniformity of the atmosphere through which the wave propagates, and work action between the wave and a flow.  In addition, the detected intensity will depend on the properties of the observing instrument.  If coronal seismology is to be a useful diagnostic tool, then we must isolate these different influences to determine which are the most important.

For a line-of-sight passing orthogonally through the axis of a coronal loop, the observed intensity of the optically-thin plasma is 
\begin{equation}
I(s) = n^2 G(T) A^{1/2} , \label{eq:9}
\end{equation}
where $n$ is the electron number density, $A$ is the cross-sectional area, $G(T)$ is the instrument response function describing the temperature-dependent sensitivity of the observing instrument, and $s$ measures position along the loop.  It is assumed here that the cross section is circular, as suggested by both observational \citep{k2000} and theoretical \citep{kan2000} considerations, and that the plasma is uniform over the cross section.  See \cite{ons00} for a discussion of the effects of cross-field gradients. 

An acoustic wave propagating along the loop will perturb the density and temperature and therefore the intensity.  The cross section will be nearly unaffected, however, because $\beta = 8\pi P / B^2$ is very small and the magnetic field behaves like a rigid tube.  Differentiating Equation \ref{eq:9} gives
\begin{equation} 
\delta I = 2 n G(T) A^{1/2} \delta n + n^2 A^{1/2} \frac{dG}{dT} \delta T , \label{eq:11}
\end{equation}
where $\delta I$, $\delta n$, and $\delta T$ are the intensity, density, and temperature perturbation amplitudes, respectively.  As discussed in the Appendix, the second term on the right-hand-side is expected to be considerably smaller than the first term for waves observed by {\it TRACE}.  We therefore drop the second term to obtain the approximate expression
\begin{equation}
\delta I \approx 2 n G(T) A^{1/2} \delta n . \label{eq:12}
\end{equation} 
It is interesting that the temperature perturbation has no direct effect on $\delta I$ even though it is crucially important for thermal conduction damping.

In the absence of dissipation, the energy density of an acoustic wave satisfies \begin{equation} 
\varepsilon \propto \frac{C_s}{A \left(v + C_s\right)^2} , \label{eq:16} 
\end{equation}
where $v$ is the velocity of a flow that may be present in the loop  (e.g., \citealt{j77}, Equation 19).  This form accounts for the exchange of work energy between the wave and the flow.  It reduces to the familiar expression given in Equation \ref{eq:18} when $v = 0$.  We can add the effect of dissipation by multiplying the right-hand-side of Equation \ref{eq:16} by $e^{-s/H_d}$, where 
\begin{equation} 
H_{d} \equiv - {\varepsilon} \left(\frac{d \varepsilon}{d s}\right)^{-1}  \label{eq:h_ddef}
\end{equation}
is the dissipation scale length for the wave energy. 
A negative sign is used so that $H_d$ will be positive for waves propagating in the direction of increasing $s$, to follow usual convention.  We refer to $H_d$ as the ``damping length" and stress that it refers exclusively to energy dissipation by direct plasma heating.  Energy loss by work is not included in $H_d$.

Comparing the modified form of Equation \ref{eq:16} with Equation \ref{eq:19}, we find that the velocity perturbation amplitude is given by
\begin{equation} 
\delta v \propto \frac{1}{C_s + v} \left( \frac{C_s}{\rho A} \right)^{\frac{1}{2}} e^{-\frac{s}{2H_d}} . \label{eq:20} 
\end{equation} 
This is in agreement with the findings of \cite{dh04} for the case of uniform temperature and no flow.
From the continuity equation, the velocity and density perturbations are related according to
\begin{equation} 
\frac {\delta v} {C_s} = \frac {\delta n} {n} . \label{eq:21}
\end{equation} 
Combining Equations \ref{eq:12}, \ref{eq:20}, and \ref{eq:21}, we obtain
\begin{equation}
\delta I \propto \frac{G(T) n^{3/2}} {\left(C_s + v\right) C_s^{1/2}}
e^{-\frac{s}{2 H_d}} . \label{eq:21b}
\end{equation}
Noting that $C_s \propto T^{1/2}$ and $n \propto P/T$, and defining the scale length of any quantity $f$ to be $H_f \equiv - f / (df/ds)$, we finally arrive at an expression for the scale length of the intensity perturbation:
\begin{equation}
H_{\delta I} \approx \left[ \frac{3}{2} H_P^{-1} + \left( \frac{T}{\Upsilon_G} - \frac{7}{4} - \frac{1}{2} \frac{1}{1+M} \right) H_T^{-1} - \frac{M}{1+M} H_v^{-1} + \frac{1}{2} H_d^{-1} \right]^{-1} , \label{eq:22}
\end{equation}
where $M \equiv v/C_s$ is the Mach number, and 
\begin{equation} 
\Upsilon_G \equiv G(T) \left(\frac {dG}{dT}\right)^{-1}  \label{eq:15}
\end{equation}
is the scale temperature of the instrument response function.  
Equation \ref{eq:22} is very useful because it reveals how the different properties of the loop affect the intensity perturbation.  It shows that the characteristic distance over which $\delta I$ varies depends directly on the characteristic distances over which the loop parameters vary.  The dependence is linear in the sense that $H_{\delta I}$ is proportional to the parameter scale length when only one parameter is non-constant.  For example, $H_{\delta I} = (2/3) H_P$ with pressure stratification alone.

The terms with the smallest scale lengths in Equation \ref{eq:22} have the biggest affect on $H_{\delta I}$.  Furthermore, the smaller a scale length is, the smaller it makes $H_{\delta I}$.  Consider the case of a static loop ($v = 0$).   If we treat each term separately, then $H_{\delta I} \propto (2/3) H_P$, $(\Upsilon_G/T) H_T$, $(-4/9) H_T$, or $2 H_d$.  We indicate two separate temperature terms because one depends on the observing instrument and the other does not.  The coefficients of the terms are all of order unity\footnote{$\Upsilon_G/T \sim 1/3$ for spectral line observations and can be much larger  for broad-band observations.}, so the relative importance of the terms is determined primarily by the relative sizes of the scale lengths.

The presence of a flow generally has minimal impact on $H_{\delta I}$.  The $(1+M) H_v / M$ term tends to be large (i.e., not important) because velocities tend to be small where velocity gradients are large, and {\it vice versa}.  
The change to the coefficient of the instrument-independent $H_T$ term is minimal, with its magnitude increasing only from 4/9 to 1/2 for $M = 1$.

Since the terms in Equation \ref{eq:22} can be either positive or negative, the intensity perturbation can either decrease or increase as the wave propagates.  Upwardly propagating waves generally experience decreasing pressures and increasing temperatures, and such conditions cause the intensity perturbation to decrease with height.  This may seem inconsistent with our demonstration in the Introduction that the velocity perturbation increases with height under normal conditions.  In fact, there is no contradiction.  Intensity and velocity perturbations are not expected to behave in the same way, which is a subtle but important point that is sometimes overlooked.  From Equations \ref{eq:deltav}, \ref{eq:12}, and \ref{eq:21}, we have that $\delta v \propto T^{1/4} P^{-1/2}$, but that $\delta I \propto T^{-9/4} P^{3/2}$, so the velocity and intensity perturbations depend on temperature and pressure in an opposite sense. 

It is interesting that $H_{\delta I}$ has no direct dependence on the cross-sectional area of the loop.  This is only true when the line-of-sight is orthogonal to the loop axis, as we have assumed, because then the area dependence of Equation \ref{eq:9} exactly cancels the area dependence of Equation \ref{eq:16}.  There is nonetheless an indirect dependence on the cross section due to its effect on the temperature and pressure structure of the loop atmosphere.  The velocity perturbation, unlike the intensity perturbation, does depend directly on the cross section, as shown in Equation \ref{eq:20} and demonstrated in \cite{dh04}.

In summary, the evolution of the intensity perturbation of a propagating acoustic wave depends primarily on the pressure and temperature structure of the loop and the wave energy damping length.  The relative importance of these different factors depends on the details of the loop atmosphere.  In the next section, we evaluate the separate terms in Equation \ref{eq:22} for three specific equilibrium loop models.

\section{Equilibrium Loop Models}

In the solar corona, where the electrical conductivity is very large and $\beta$ is very small, the plasma is confined by the magnetic field and its evolution is described by the one-dimensional conservation equations of mass, momentum and energy: 
\begin{equation}
{ { \partial \rho }\over{ \partial t  } } + 
{{1}\over{A}}{{ \partial  }\over{ \partial s }} \left( A v \rho 
\right) = 0 ,
\label{eq:1}
\end{equation}
\begin{equation} 
{ { \partial (v \rho)  }\over{ \partial t  } }  +  
{{1}\over{A}}{{ \partial ( A  v^{2} \rho ) }\over{ \partial s }} 
 + {{ \partial P  }\over{ \partial s }}
= \rho g_{\parallel}  + \rho a ,
\label{eq:2}
\end{equation} 
\begin{equation}
{ { \partial E  }\over{ \partial t  } }  + 
 {{1}\over{A}}{{ \partial \left[ A (E+P)v \right] }\over{ \partial s }}  = 
\rho v g_{\parallel}   + {{1}\over{A}} {{ \partial }\over{ \partial s }} 
\left( A  \kappa_{0} T^{5/2} {{ \partial T }\over{ \partial s }} \right) 
-n^{2} \Lambda(T) +  Q ,
\label{eq:3}
\end{equation}
where
\begin{equation} 
E= {{1}\over{2}} \rho v^{2} + {{P}\over{ \gamma - 1 }}
\label{eq:4} .
\end{equation}
As before, $T$, $P$, $\rho$, $v$, and $A$ are the temperature, pressure, mass density, velocity, and cross-sectional area, respectively; $s$ is distance along the loop from the left base of the model; $g_{\parallel}(s)$ is the component of gravity parallel to the loop axis; $Q(s)$ is the volumetric heating rate; $\Lambda(T)$ is the optically-thin radiation loss function; $\kappa_{0} = 10^{-6}$ is the coefficient of thermal conduction; and $\gamma = 5/3$ is the ratio of the specific heats.  We assume a fully ionized hydrogen plasma, so $\rho = m_p n$ and $P = 2 k_B n T$, where $n$ is the electron number density and $k_B$ is Boltzmann's constant.  We do not include compressive viscosity, because it is generally small under the conditions we will investigate \citep{pks94a,dh03}.  Finally, $a=a(s,t)$ is a spatially localized acceleration that we impose to generate acoustic waves.  It is set to zero for the equilibrium solutions.

The geometry of the loop is defined by the 
specific form of $g_{\parallel}(s)$. We adopt a semicircular shape for the coronal part, so that 
\begin{equation} 
g_{\parallel}(s) = g_{\odot} 
\left[ \frac{R_\odot}{R_\odot + h(s)} \right]^{2} 
\cos \: \left[\theta (s) \right] , \label{eq:5} 
\end{equation}
where $h(s)$ is the height above the solar surface, $g_{\odot}$ is the gravitational acceleration 
at the surface,   $R_{\odot}$ is the solar radius, and  $ \theta (s) $ is the angular position 
along the semicircle. This form accounts for the decrease in gravity with height, which can be important for high-arching loops.  We choose a coronal length $L = 3\times10^{10}$ cm.  Attached to the corona at both ends are $6\times10^9$ cm long chromospheric sections, giving a total length for the model of $4.2\times10^{10}$ cm.

The optically thin radiative loss function, $\Lambda (T)$, is a piecewise 
continuous power law as described in \cite{kc2001}.  It is based on the atomic physics calculations of J. Raymond (private communication) and uses abundances twice as large as the coronal values given by \cite{m85}.  The loss function drops precipitously to zero between $3.0\times10^4$ and $2.95\times10^4$ K in order to maintain a nearly constant chromospheric temperature within this range.  The detailed physics of the chromosphere involves optically-thick radiative transfer and is not treated, but our model accurately reproduces the energetic and dynamic coupling between the corona and chromosphere which is vital to a realistic simulation.

Although there is evidence that many coronal loops are inherently time dependent (e.g., \citealt{ck97}), we limit ourselves in this study to equilibrium loops.  We consider two static models and one model with steady end-to-end flow.  Static equilibrium requires that the heating function, $Q(s)$, be symmetric with respect to the loop midpoint.  In order to produce as flat a temperature profile as possible, as suggested by {\it TRACE} observations \citep{letal99,ana00}, we adopt a heating function that decreases exponentially with height in both legs.  In the ``left" leg,
\begin{equation} 
Q(s) = Q_0 \:  \exp  \left( - {{s - s_0}\over{ H_Q }} \right) , \label{eq:6}
\end{equation}
where $H_Q = 7.5\times10^9$ cm is the heating scale length and $Q_0 = 4.6\times 10^{-6}$ erg cm$^{-3}$ s$^{-1}$ is the heating rate at the left footpoint ($s_0 = 6\times10^9$ cm).  The heating function is a mirror image in the ``right" leg.  We have set the magnitude of the heating to produce a peak temperature of approximately 1.2 MK, characteristic of {\it TRACE} loops. 
 
To obtain a steady flow solution, we follow \cite{pkm03} and apply an exponential heating function over the entire loop, without mirror reflection.  The heating rate decreases from the left base, past the loop midpoint, all the way to the right base.  In this case we use a scalelength $H_Q = L/10 = 3\times10^9$ cm and magnitude $Q_0=6.5 \times 10^{-5}$ erg cm$^{-3}$ s$^{-1}$, again chosen to produce a peak temperature near 1.2 MK.  The extreme heating asymmetry drives a strong left-to-right flow in the loop.

\begin{figure}
\plotone{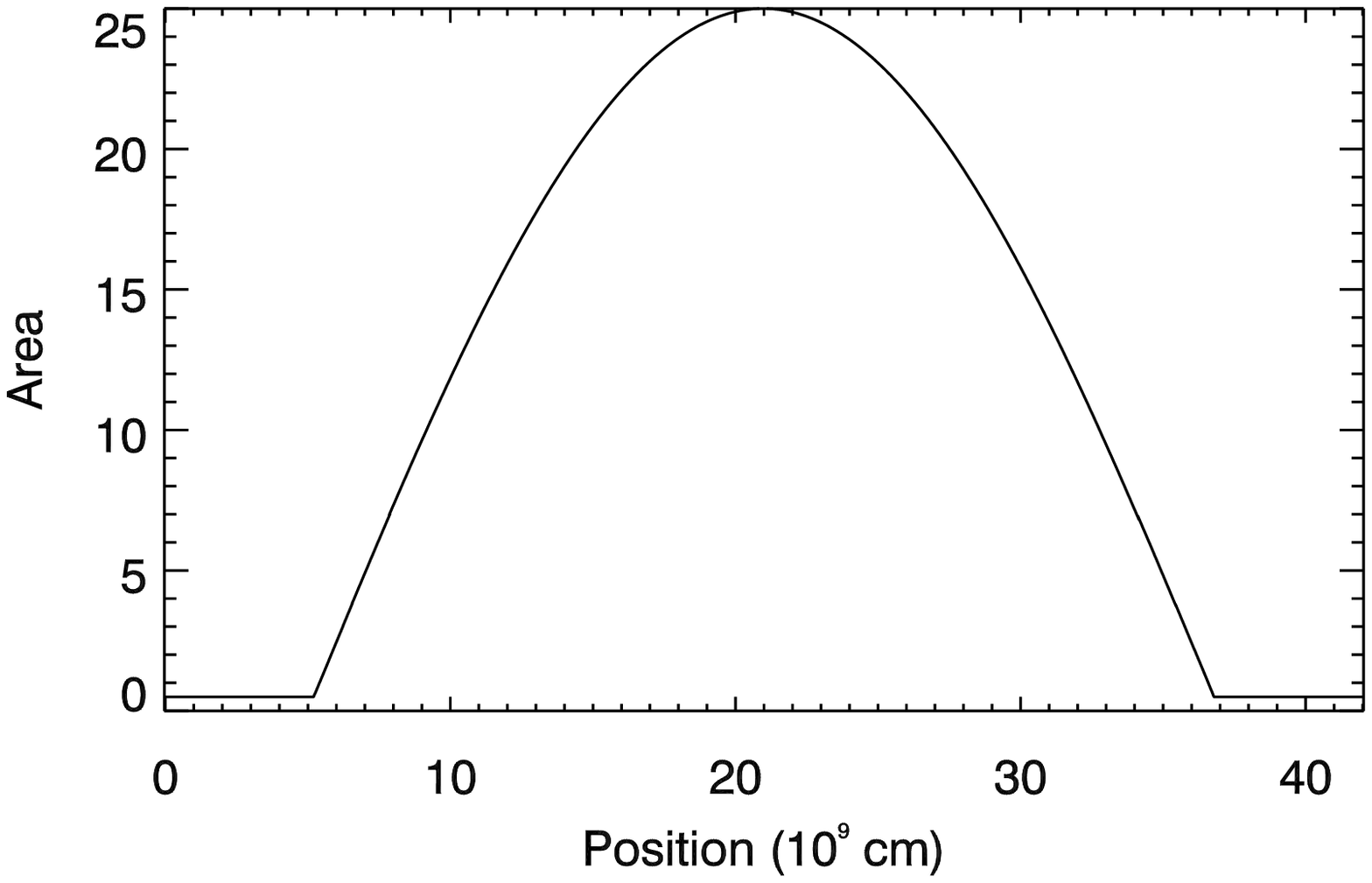}
\caption{Cross-sectional area versus position along the loop in Case 3.
\label{fig:avss}}
\end{figure}

Coronal loops that are visible along their entire length tend to have constant cross sections \citep{kea1992,k2000,wk2000,lf2004}.  Acoustic waves are typically observed in much longer loops, for which only the lower legs are visible.  There is some evidence that these loops expand with height.  \cite{dhiw2002a} measured an average expansion rate of $0.28 \pm 0.16$ (width change per unit length).  The average width of the loops is $8.1 \pm 2.8 \times 10^8$ cm, so the expansion rate can be expressed as 3.5\% per $10^8$ cm.  In one of our models, we assume a cross section that expands according to 
\begin{equation} 
A(s)  = A_{mid} \: \cos \left[ {{\chi \pi}\over{2}} 
{{(s_{mid}-s)}\over{s_{mid}}} \right], 
\end{equation}
where $A_{mid}$ is the area at the loop midpoint ($s_{mid} = 2.1 \times 10^{10}$ cm) and we set $\chi = 1.3$ to give a width expansion rate of 4.3\% per $10^8$ cm at the base of the corona (area expansion rate of 8.7\%).
The cross section is taken to be constant throughout most of the chromosphere. Figure \ref{fig:avss} shows the variation of area with position in the loop.

Equations (\ref{eq:1}-\ref{eq:4}) are solved numerically using a code we call 
Adaptively Refined Godunov Solver (ARGOS). As described in \cite{amsk99}, it utilizes a  
second order Riemann type solver \citep{v79} to treat the one-dimensional equations in the absence of sources. 
The sources are accounted for using an operator splitting method in second order. An important feature of ARGOS is  
the parallel adaptive mesh refinement package PARAMESH, which has been set to refine 
or derefine the local grid based upon density variations.  Rigid wall boundary conditions are imposed at the ends of the model, which are many gravitional scale heights deep in the chromosphere.
To obtain an equilibrium solution, we begin with a reasonable guess of the loop atmosphere based on well-known scaling laws (e.g., \citealt{rtv78}), and allow the solution to relax asymptotically to a steady state.  

\begin{figure}
\plotone{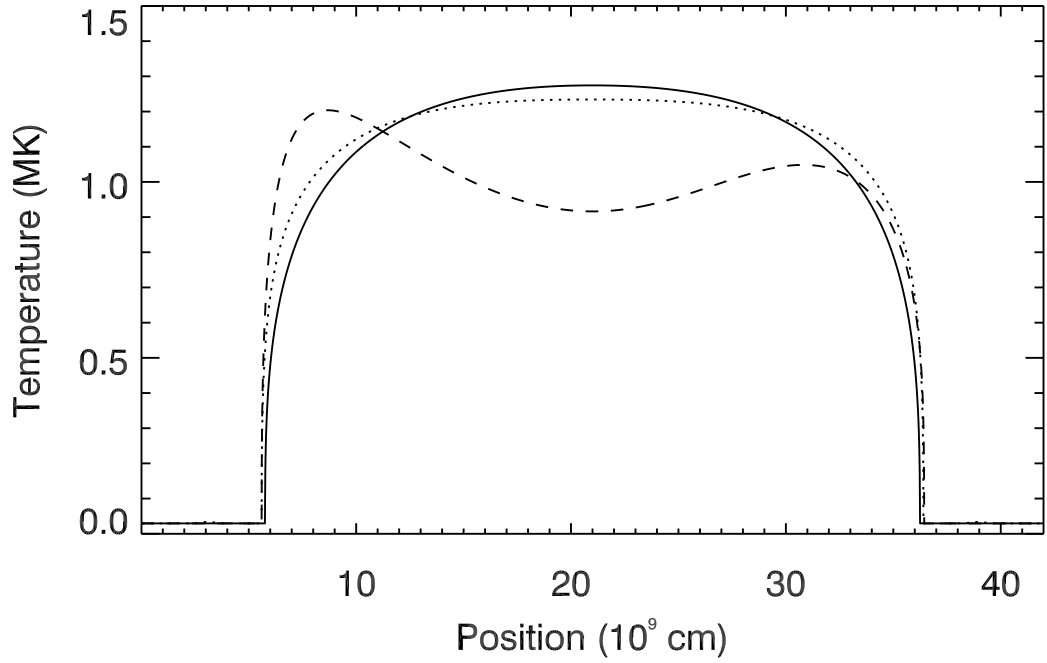}
\caption{Temperature versus position along the loop in Cases 1 (solid), 2 (dashed), and 3 (dotted).
\label{fig:tvss}}
\end{figure}

\begin{figure}
\plotone{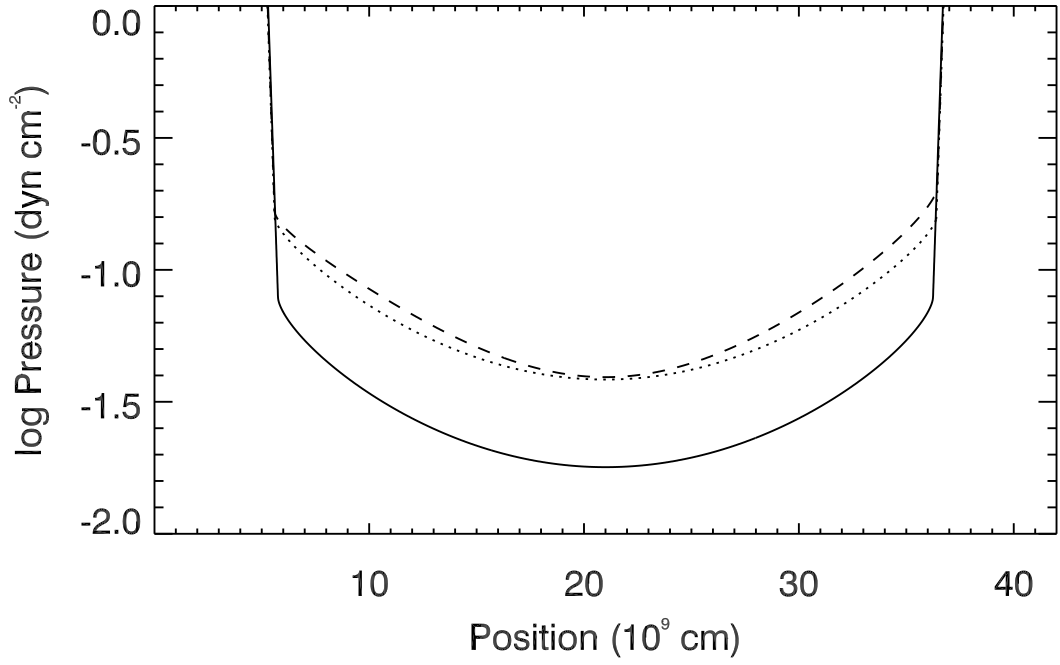}
\caption{Pressure versus position along the loop in Cases 1 (solid), 2 (dashed), and 3 (dotted).
\label{fig:pvss}}
\end{figure}

\begin{figure}
\plotone{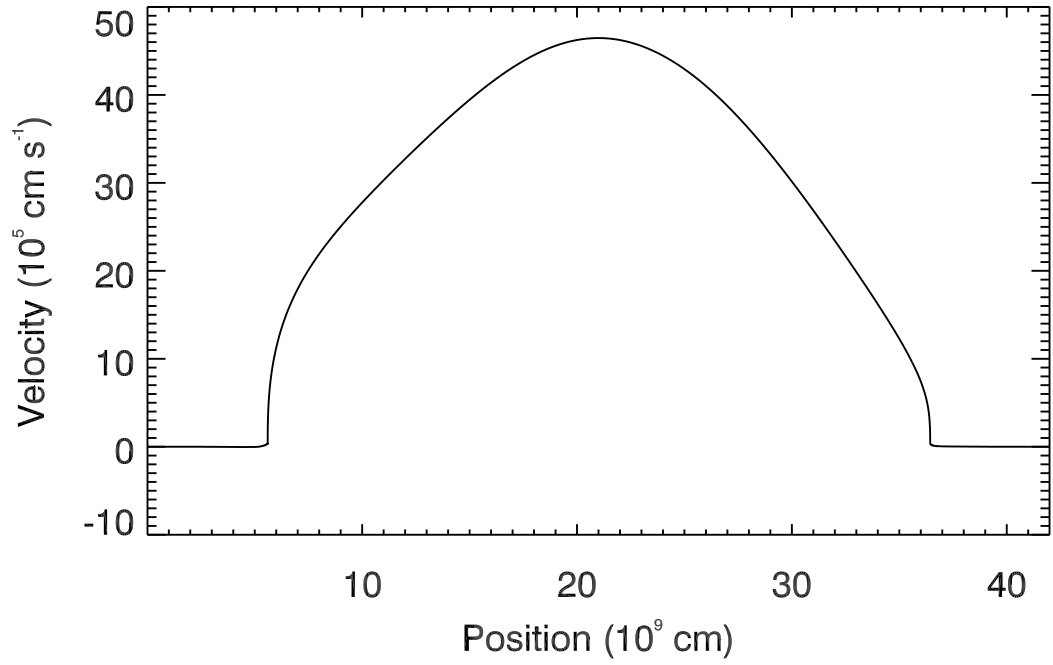}
\caption{Velocity versus position along the loop in Case 2.
\label{fig:vvss}}
\end{figure}

We consider three different equilibrium loops:             
Case 1 is a constant cross section loop with symmetric heating (static equilibrium); Case 2 is a constant cross section loop with asymmetric heating (steady flow equilibrium); and Case 3 an expanding cross section loop with symmetric heating (static equilibrium).  
Figures \ref{fig:tvss} and \ref{fig:pvss} show the profiles of temperature and pressure, respectively, for all three cases.  Figure \ref{fig:vvss} shows the profile of velocity for Case 2.  The maximum velocity is 46 km s$^{-1}$ and corresponds to a Mach number of 0.29.  There are clear differences among the temperature and pressure profiles for the three cases.  In particular, the temperature profile in Case 2 has an inversion that is not present in Cases 1 and 3.  The physical explanation for this behavior is given in \cite{pkm03}.  Differences between Cases 1 and 3 can be understood in terms of changes to the energy balance associated with the constriction of the loop footpoints.  For Case 3, a greater fraction of the coronal heating energy must be radiated from the upper parts of the loop and less conducted down to the lower parts, so the density is higher and the temperature profile flatter.

Figures \ref{fig:hvss1}, \ref{fig:hvss2}, and \ref{fig:hvss3} indicate the values of the terms in Equation \ref{eq:22} as a function of position along loop for Cases 1, 2, and 3, respectively.  Only the left half of the loop is shown.  The $H_P$, $H_T$, and $H_v$ terms were obtained directly from the equilibrium models.  Of these, the pressure term is most important (i.e., has the smallest magnitude) throughout most of the corona.  Pressure is gravitationally stratified and therefore $H_P$ increases toward the apex as the loop bends to become more and more horizontal.  The temperature terms are important in the lower corona, and they dominate in the transition region where the gradients are very steep.  The interesting shape of the instrument-dependent $H_T$ curve is due to the form of $G(T)$.  We have used a $G(T)$ corresponding to the {\it TRACE} 171 channel, which peaks at $T = 0.95$ MK.  The curve is vertical at the position where this temperature occurs, because $\Upsilon_G \propto (dG/dT)^{-1}$ switches discontinuously from large positive to large negative values.  The $H_T$ terms are negative in the plotted portion of Case 2.  This is because of the temperature inversion.
The $H_v$ term is insignificant.  

\section{Wave Simulations}

In principle, we can use linear theory to estimate the value of the damping length from the equilibrium models and then solve for $H_{\delta I}$ in Equation \ref{eq:22}.  Following \cite[Eqn. 4.1.2f]{pks94a}, we have that
\begin{equation}
H_d^* \approx 146 \frac {n \tau^2} {T} , \label{eq:23}
\end{equation}
where the asterisk indicates that the expression applies to adiabatic waves propagating in a uniform medium.  For Case 1, $H_d^* \approx 9.7\times10^8$ cm at position $s = 1.0\times10^{10}$ cm.  We show in the Appendix, however, that waves of the observed periods are approximately isothermal under the conditions in the loop.  Furthermore, the wavelengths are not small compared to both the pressure and temperature scale lengths.  Equation \ref{eq:23} is therefore not valid.  To determine $H_{\delta I}$ and $H_d$ we must perform a nonlinear simulation of wave propagation.

We generate waves in our loops using a sinusoidal acceleration that is localized just above the left footpoint:  
\begin{equation} 
a(s,t) = a_0 \: \exp \left[ - \left( {{s-s_{1}  }
\over{ \sigma  }} \right)^{2} \right] 
\sin \left(  \frac{2\pi t}{\tau} \right) . \label{eq:8}
\end{equation}
The Gaussian spatial profile has a 1/{\it e} halfwidth $\sigma = 10^7$ cm and is centered at $s_1 = 6.2\times10^9$ cm for Cases 1 and 3, and $s_1 = 8.5\times10^9$ cm for Case 2.  The temperature at these locations is 0.55, 1.20, and 0.73 MK for Cases 1, 2, and 3, respectively.  We choose an oscillation period $\tau = 250$ s, similar to those observed by {\it TRACE}, and an acceleration magnitude $a_0$ that produces a wave amplitude $\approx 4 \%$ ($\delta I \approx 8 \%$), about twice as large as the average observed by {\it TRACE}.

\begin{figure}
\plotone{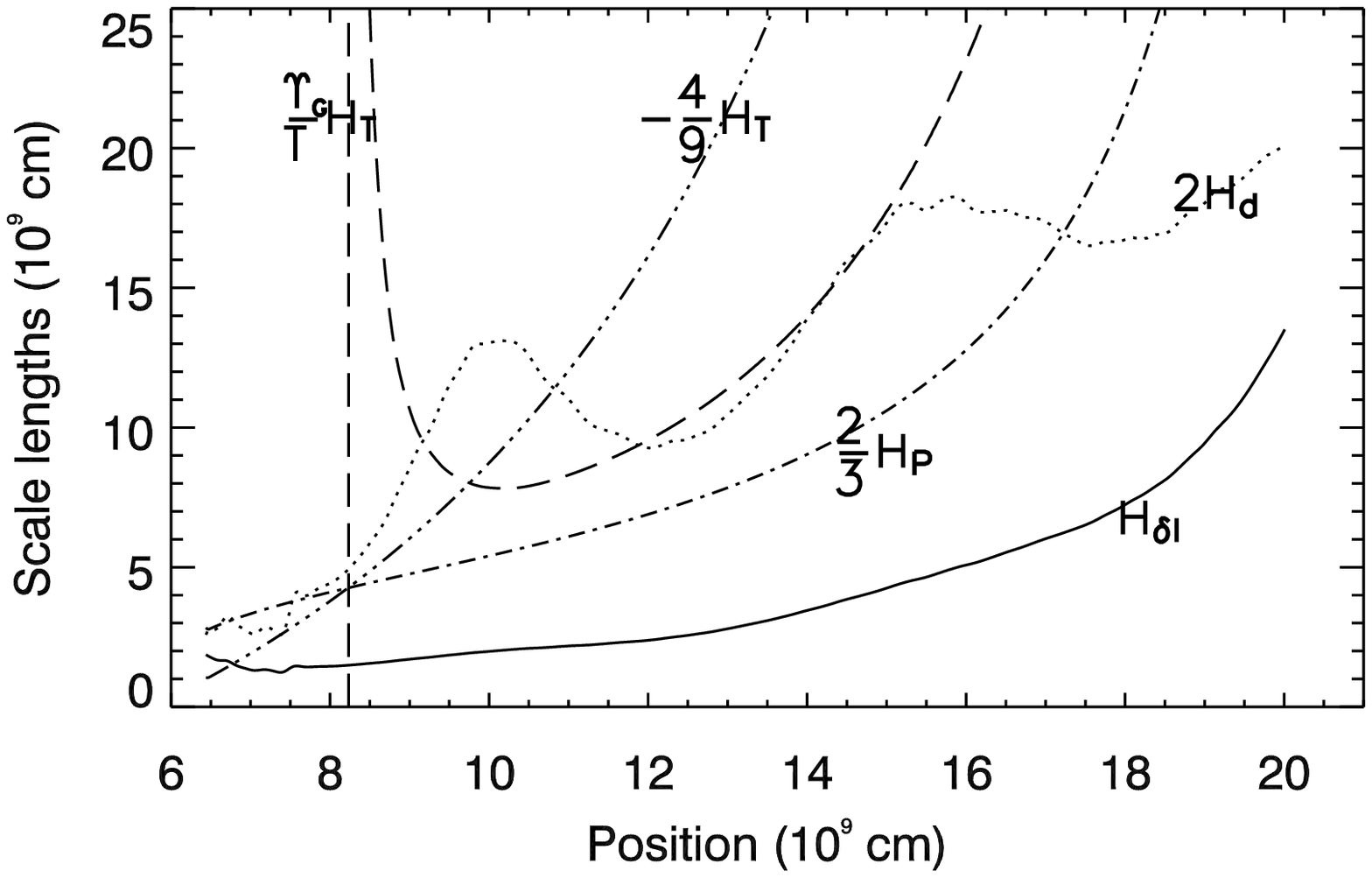}
\caption{Values of the terms in Equation \ref{eq:22} versus position along the loop for Case 1:  $H_P$ and $H_T$ terms are from the equilibrium model; $H_{\delta I}$ is from the wave simulations; and $2H_d$ is from the equation.
\label{fig:hvss1}}
\end{figure}

\begin{figure}
\plottwo{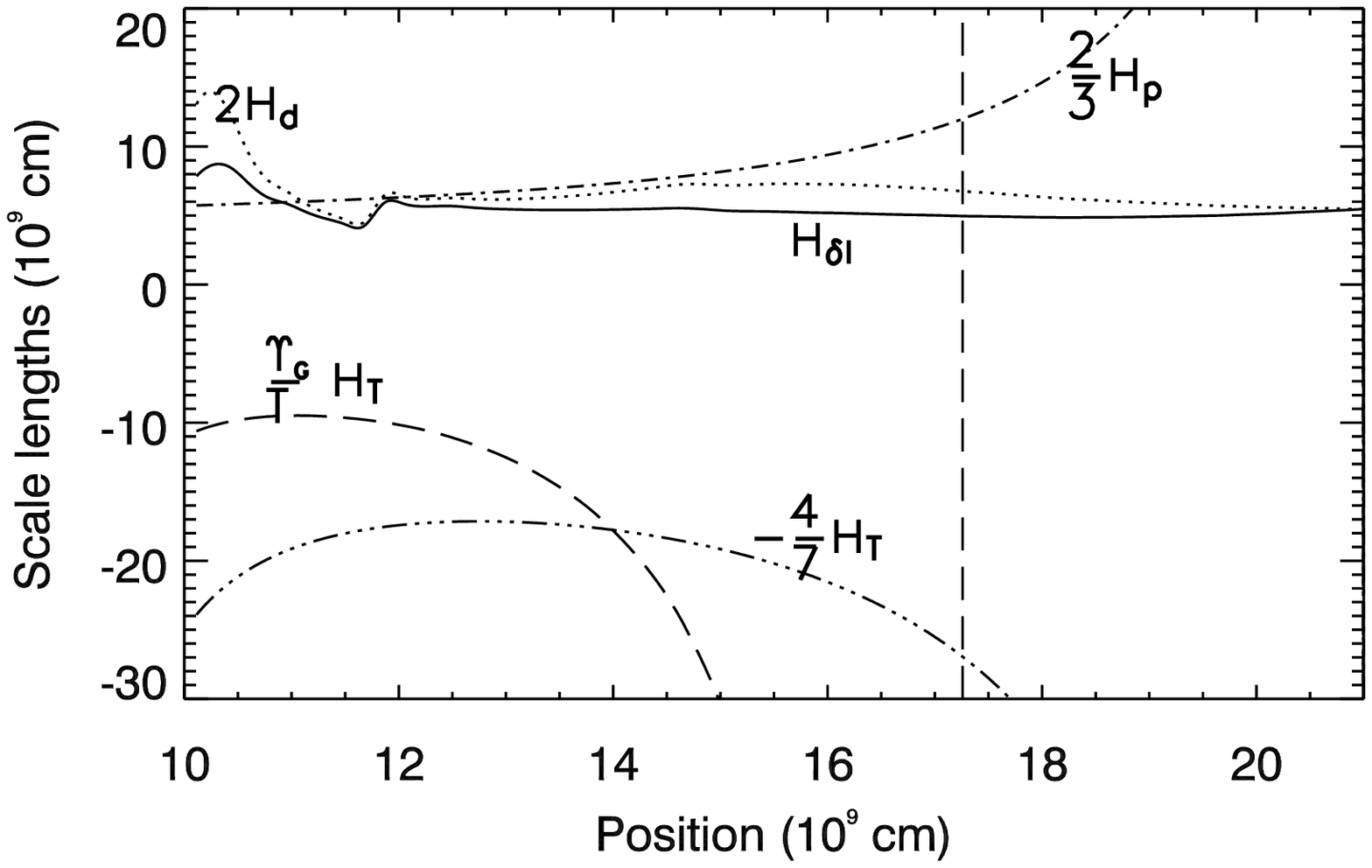}{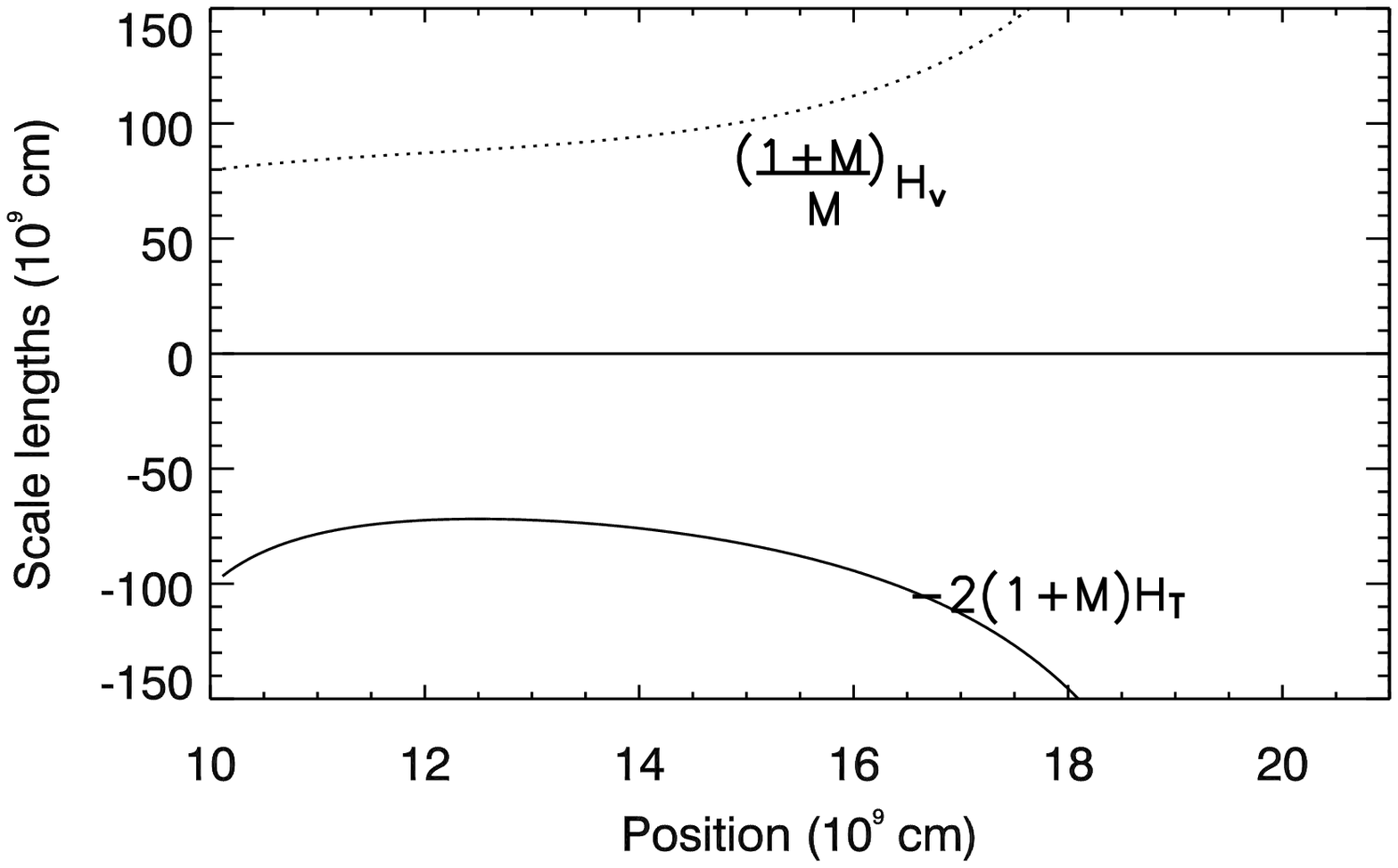}
\caption{Values of the terms in Equation \ref{eq:22} versus position along the loop for Case 2:  $H_P$, $H_T$, and $H_v$ terms are from the equilibrium model; $H_{\delta I}$ is from the wave simulations; and $2H_d$ is from the equation.
\label{fig:hvss2}}
\end{figure}

\begin{figure}
\plotone{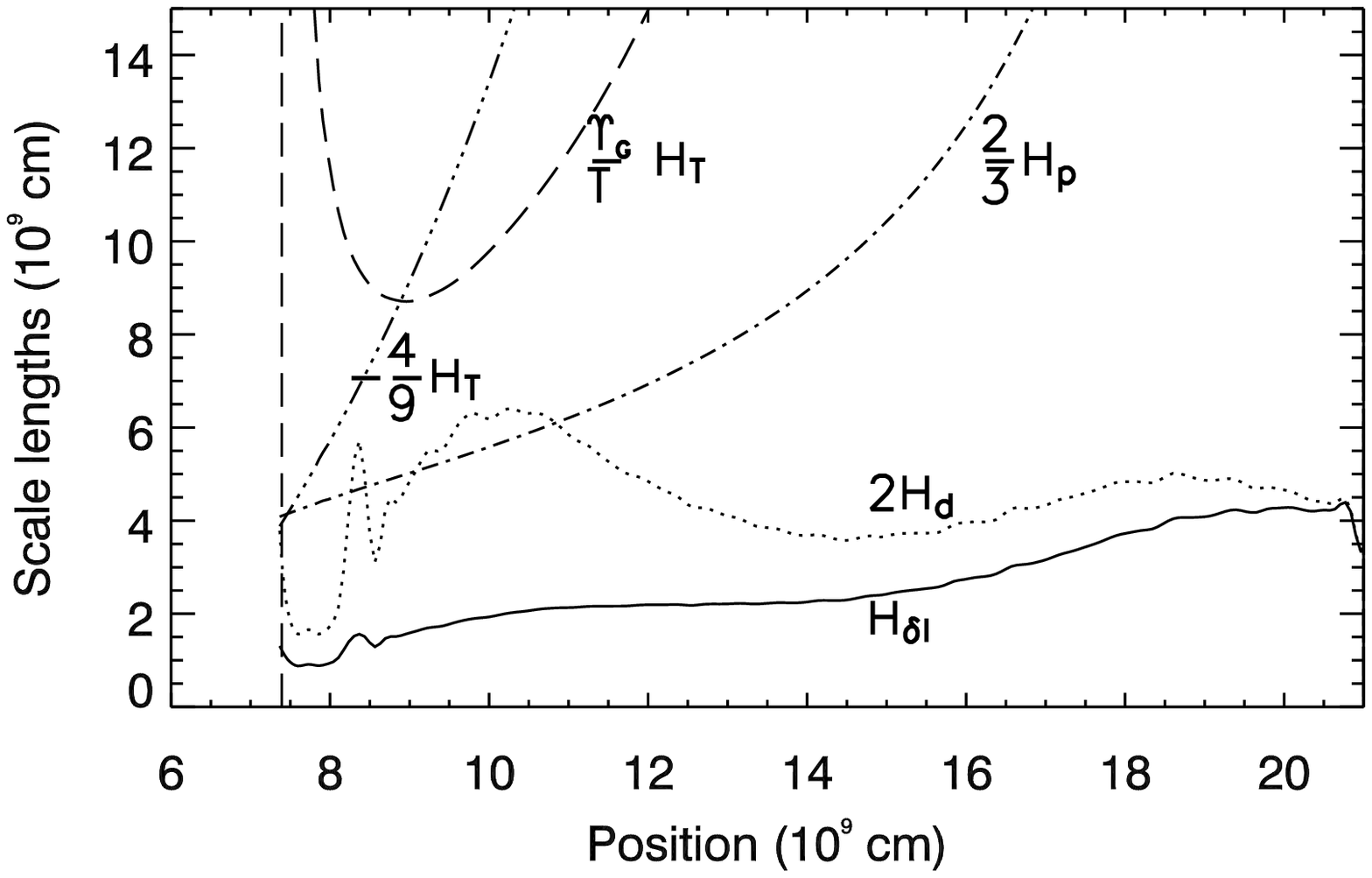}
\caption{Values of the terms in Equation \ref{eq:22} versus position along the loop for Case 3:  $H_P$ and $H_T$ terms are from the equilibrium model; $H_{\delta I}$ is from the wave simulations; and $2H_d$ is from the equation.
\label{fig:hvss3}}
\end{figure}

\begin{figure}
\plottwo{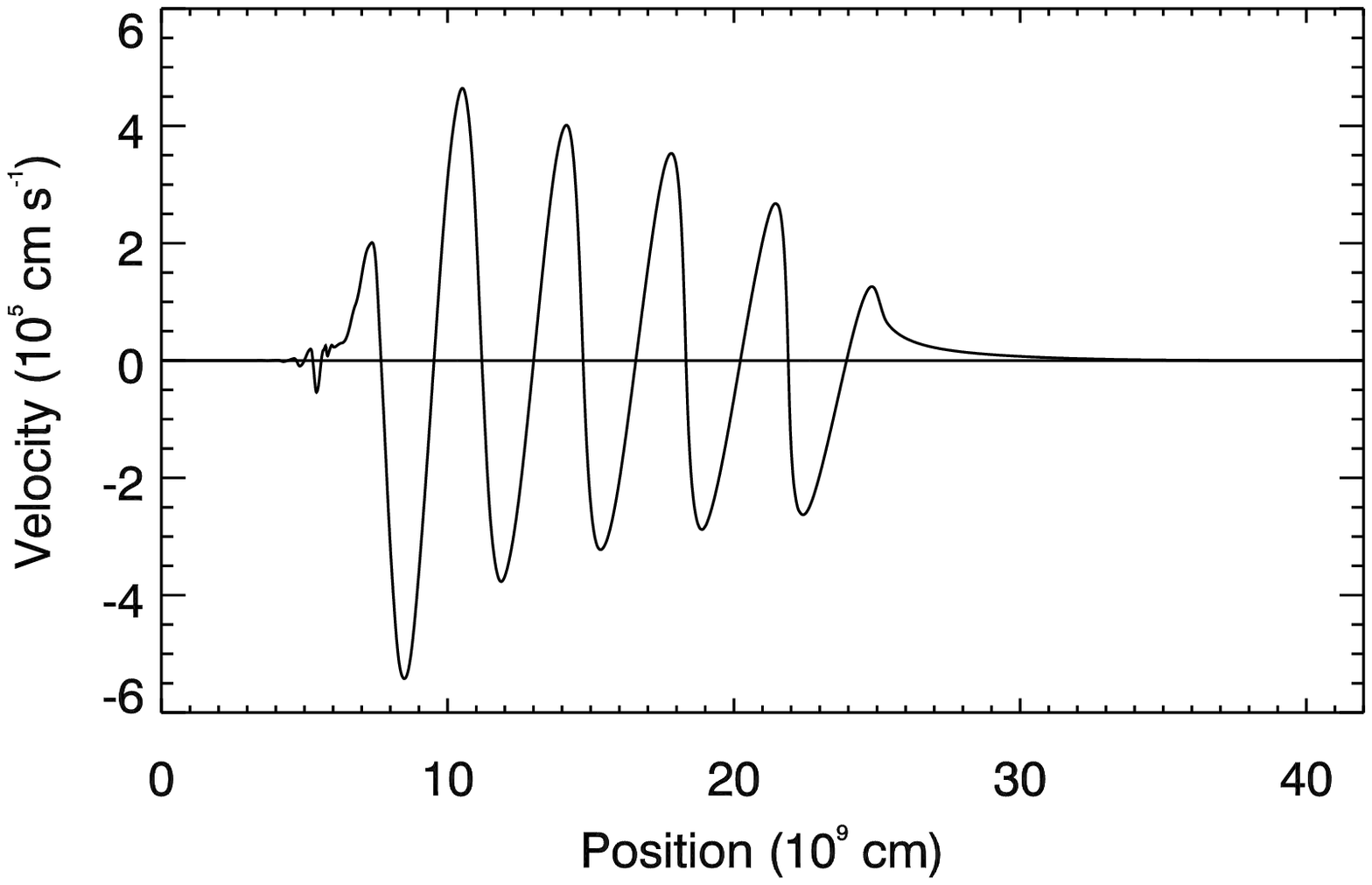}{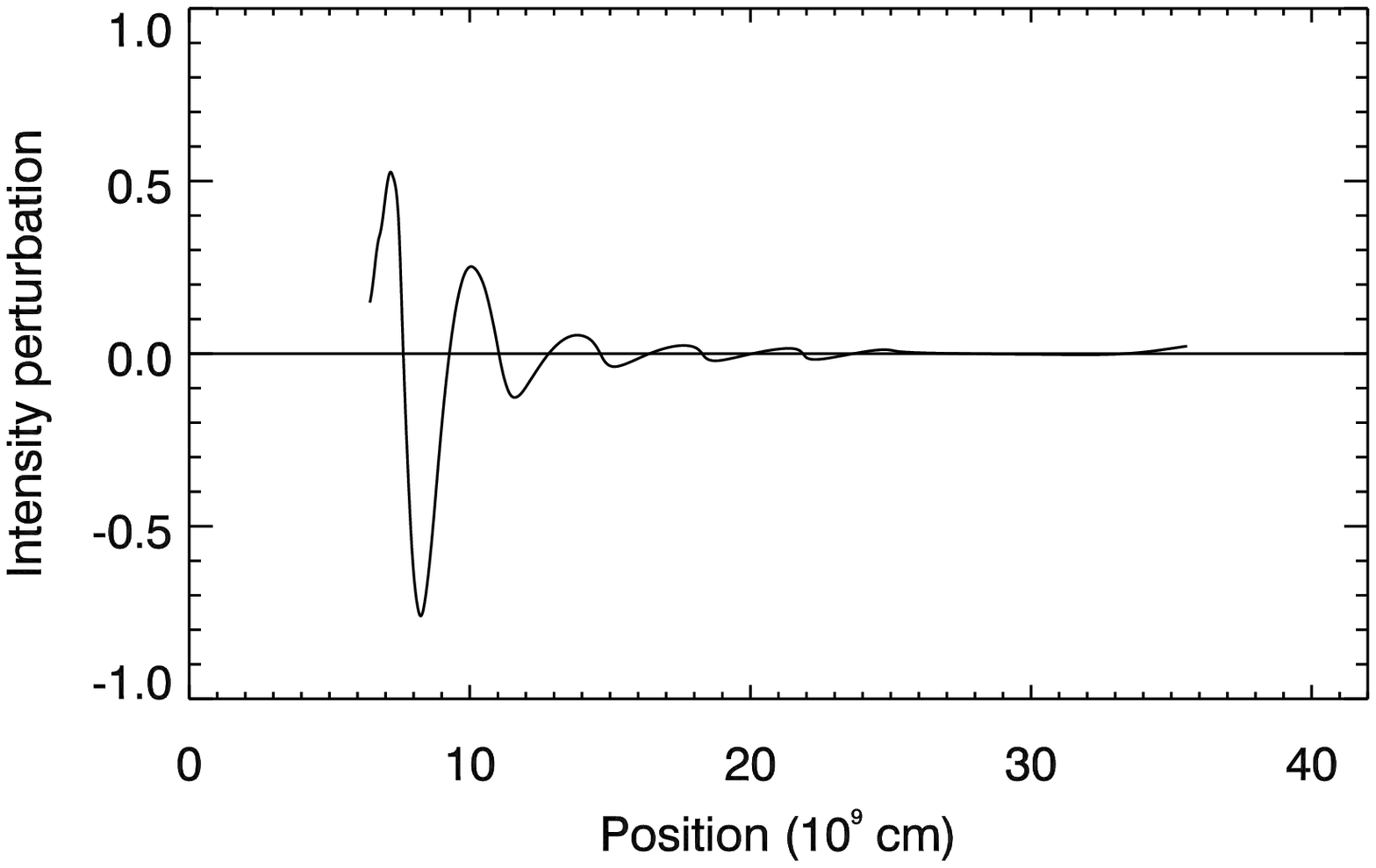}
\caption{Velocity perturbation (left) and intensity perturbation (right) versus position along the loop in Case 1.  Intensities were computed assuming observations made in the {\it TRACE} 171 channel.
\label{fig:pert}}
\end{figure}

\begin{figure}
\plotone{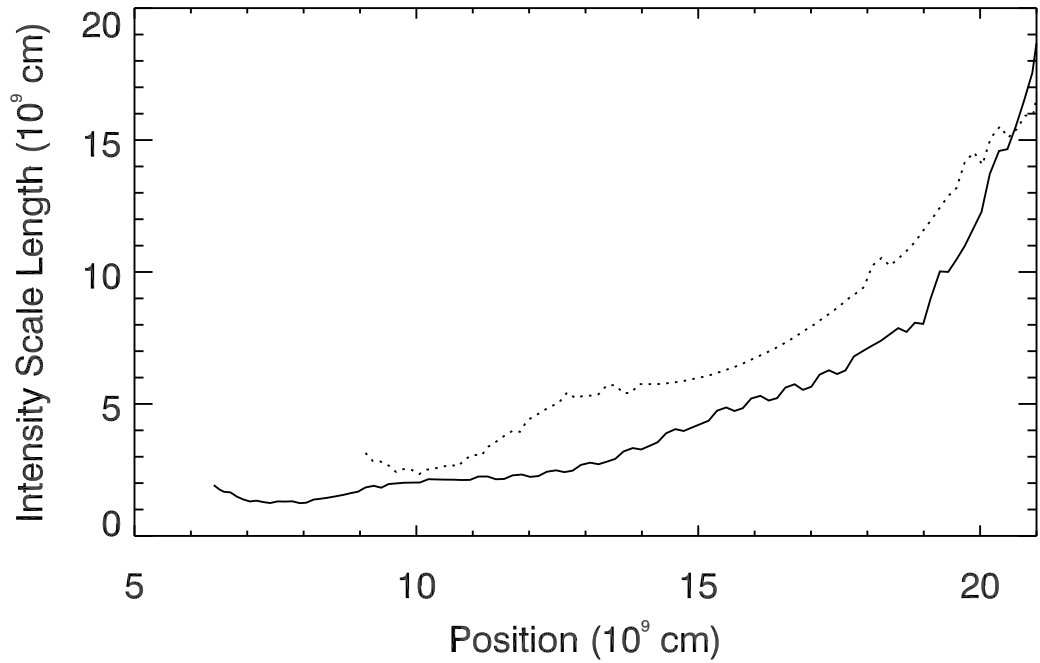}
\caption{Intensity scalelength versus position along the loop in Case 1 for simulated observations in the 171 (solid) and 195 (dotted) channels of {\it TRACE}.  Wiggles in the curves are numerical artifacts.
\label{fig:h171195}}
\end{figure}

The left panel of Figure \ref{fig:pert} shows the velocity perturbation in Case 1 at a time when six wave fronts have been generated.  The right panel shows the corresponding intensity perturbation computed from Equation \ref{eq:9} for a simulated {\it TRACE} 171 observation.  Its amplitude decreases rapidly as the wave propagates along the loop.  By following one of the wave fronts, we determine the scale length 
\begin{equation} 
H_{\delta I} \equiv - {\delta I} \left(\frac{d \delta I}{d s}\right)^{-1} . \label{eq:10}
\end{equation}
The solid curve in Figure \ref{fig:h171195} shows $H_{\delta I}$ as a function of position along the loop for the second wave front.  The other wave fronts give similar scale lengths.  The dashed curve in the figure shows $H_{\delta I}$ for the 195 channel.  Differences are due entirely to differences in temperature responses of the two channels.  The wiggles in the curves are not physical, but arise from small inaccuracies in determining the position of the wave front  (the position of maximum perturbation) on the finite numerical grid.  A smoothed version of the 171 curve is shown as the solid curve in Figure \ref{fig:hvss1}.  Corresponding curves for Cases 2 and 3 are shown Figures \ref{fig:hvss2} and \ref{fig:hvss3}.  We see that $H_{\delta I}$ is typically between 1 and 6$\times 10^9$ cm and somewhat larger near the top of the loop in Case 1.  This agrees very well with the actual values observed by {\it TRACE}, as discussed in the next section.

The dotted curves in Figures \ref{fig:hvss1}, \ref{fig:hvss2}, and \ref{fig:hvss3} indicate $2H_d$ as computed from Equation \ref{eq:22} using $H_{\delta I}$ from the simulations.  The damping term is of comparable importance to the pressure and temperature terms in the low corona.  It is of significantly greater importance in the upper corona, although waves there are difficult to detect because of their reduced intensity (Fig. \ref{fig:pert}).  We note that $H_d$ is larger than $H_d^*$.  This is expected since (approximately) isothermal waves dissipate more slowly than adiabatic waves \citep{dh03}.

\section{Observations} \label{sec:obs}

Using {\it TRACE} 171 observations, \cite{dhiw2002a} studied waves propagating upward in the lower corona of long loops.  They performed a wavelet analysis on the intensity perturbations and measured a quantity they call the ``detection length."  It is defined to be the distance over which the wavelet power exceeds the 99\% confidence level with respect to photon counting noise (i.e., the distance over which the intensity perturbation is at least a $2.1 \sigma$ detection).  At greater altitudes the oscillation signal vanishes into the noise.

The relationship between the detection length and the intensity perturbation scale length is not straightforward.  Whereas $H_{\delta I}$ indicates the characteristic distance over which the intensity perturbation varies and is independent of the signal strength, the detection length depends on both the perturbation amplitude and the intensity of the unperturbed loop.  For a given $H_{\delta I}$, a large amplitude wave would be ``detectable" out to greater distances than a small amplitude wave and would therefore have a longer detection length.  

To facilitate a direct comparison with our simulations, we measure an observed intensity scale length:
\begin{equation} 
H_{\delta I}^{obs} \equiv - \left(\frac{\delta I_1 + \delta I_2}{2}\right)
   \left(\frac{\delta I_2 - \delta I_1}{s_2 - s_1}\right)^{-1} , \label{eq:H_obs}
\end{equation}
where $\delta I_1$ and $\delta I_2$ are the intensity perturbation amplitudes at positions $s_1$ and $s_2$, which are the first and last positions where the wavelet power meets the 99\% criterian.  We estimate $\delta I_1$ and $\delta I_2$ as follows.  We first compute a time sequence of the intensity at each position by averaging over a cross-sectional slice that is 2-4 pixels thick.  We then subtract a linear fit of the time sequence, i.e., we remove the unperturbed signal taking into account any fading or brightening trend that may be present.  Finally, we compute a root-mean-square (RMS) intensity from the residuals, which we claim is directly proportional to the intensity perturbation amplitude.  The constant of proportionality is unimportant, since it falls out of the ratio in Equation \ref{eq:H_obs}.

\begin{figure}
\plotone{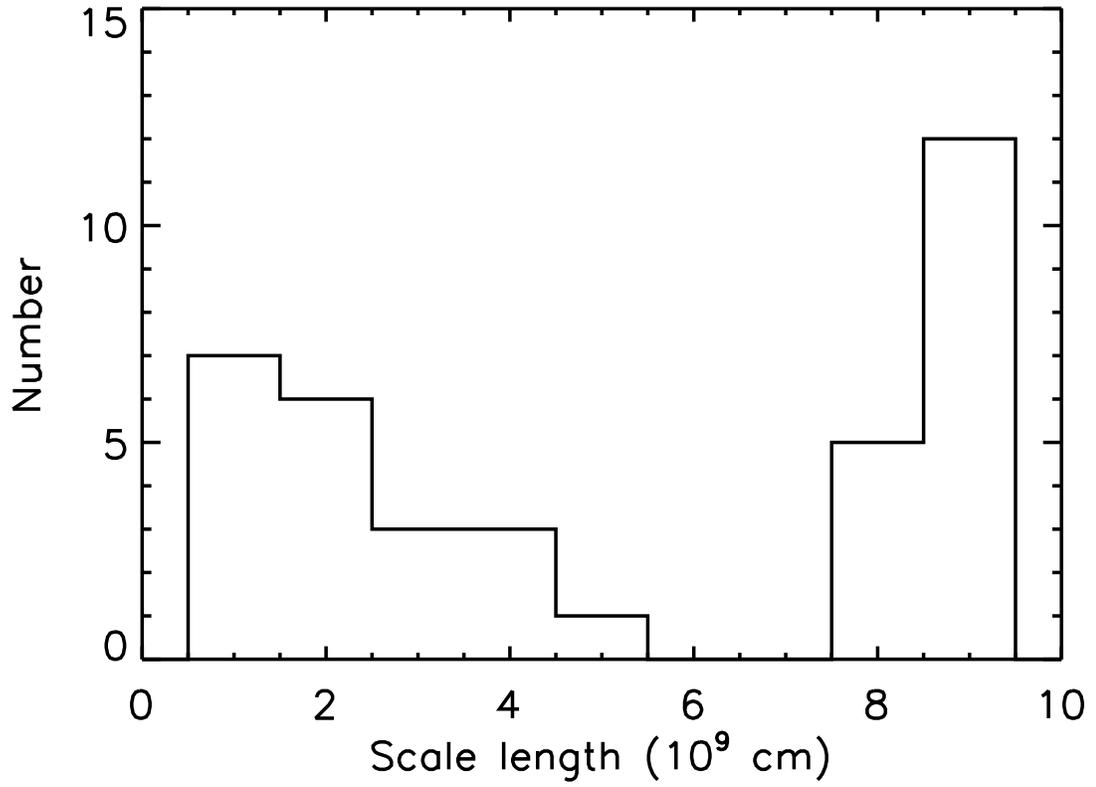}
\caption{Histogram distribution of observed intensity scale lengths.  Values greater than $8\times10^9$ cm have been assigned a value of $8\times10^9$ cm, and negative values have been assigned a value of $9\times10^9$ cm.
\label{fig:dist}}
\end{figure}

Table \ref{tbl-1} lists $\delta I_1$, $\delta I_2$, $\Delta s \equiv s_2 - s_1$ (the detection length), and $H_{\delta I}^{obs}$ for the 17 loops studied originally by \cite{dhiw2002a}.  Several of the loops were observed multiple times.  
The distribution of $H_{\delta I}^{obs}$ is shown as a histogram in 
Figure \ref{fig:dist}.  Values greater than $8\times10^9$ cm have been assigned to the $8\times10^9$ cm bin.  Nearly one-third of the scale lengths are negative.  We choose not to plot them to the left of the positive values since a more sensible physical ordering is to begin with wave amplitudes that decrease rapidly with height (small positive $H_{\delta I}^{obs}$), progress to wave amplitudes that decreases slowly with height (large positive $H_{\delta I}^{obs}$), and finally to wave amplitudes that increase with height (negative $H_{\delta I}^{obs}$).  We therefore lump the negative values together and assign them to the $9\times10^9$ cm bin.  With these changes, the median (not the mean) of the distribution is $4.35\times10^9$ cm.  The $H_{\delta I}$ from our wave simulations generally range between 1 and 6$\times 10^9$ cm, which is in excellent agreement with the observations.

The negative $H_{\delta I}^{obs}$ measurements, indicating intensity amplitudes that increase with height, are most likely a result of a temperature inversion.  The pressure term in Equation \ref{eq:22} is positive due gravitational stratification; the dissipation term is positive by definition; and the flow term is probably very small, as we have discussed.  Only the temperature terms are left to make $H_{\delta I}^{obs}$ negative.  Since $\Upsilon_G < 0$ for $T > 0.95$ MK, $H_T$ must be positive and temperature must decrease with height (i.e., there must be a temperature inversion compared to the usual situation).  We have seen that a temperature inversion is present in the steady flow equilibrium of Case 2.  The temperature gradient is not steep enough, however, to drive $H_{\delta I}$ negative in this model.  \cite{dph03} found evidence for temperature inversions in two loops of their sample for which they had both 171 and 195 observations and could apply filter ratio diagnostics.  We plan to examine other cases for which data in both channels are available.

Some readers may find it strange that $H_{\delta I}^{obs}$ could be negative at the same time that the detection length is finite.  This situation occurs whenever the unperturbed loop intensity, and hence photon counting noise and detection threshold, increase with height faster than the intensity perturbation.  The effect of increasing detection threshold with height also explains why positive $H_{\delta I}^{obs}$ values tend to be larger than or much larger than the corresponding detection lengths.

We caution that the uncertainties in our measurements are quite large.  
Following \cite{kg95}, it can be shown that the uncertainty in $H_{\delta I}^{obs}$ is given approximately by
\begin{equation} 
\Delta H_{\delta I}^{obs} \approx \frac{2 \delta I_1 \delta I_2} {\left(\delta I_1 - \delta I_2\right) \left(\delta I_1 + \delta I_2\right)} 
\left[ \left(\frac{\Delta \delta I_1}{\delta I_1}\right)^2 + \left(\frac{\Delta \delta I_2}{\delta I_2}\right)^2 \right]^{1/2} H_{\delta I}^{obs} ,
\label{eq:H_unc}
\end{equation}
where $\Delta \delta I_1 / \delta I_1$ and $\Delta \delta I_2 / \delta I_2$ are the fractional uncertainties in the intensity amplitudes.  It is difficult to determine the formal uncertainties in the intensity amplitudes, but visual inspection of the time sequences suggests that 15\% is a reasonable estimate.  With this value, the uncertainties in the scale lengths are as listed in the final column of Table 1.  It is evident that many of the scale length measurements are, by themselves, rather dubious.  However, because the measurement errors are of a random nature, so that the true scale length is sometimes underestimated and other times overestimated, and because the sample of measurements is reasonably large, we feel that the basic results discussed above are still valid.

\section{Discussion}

We have shown that the intensity perturbation of an acoustic wave propagating along a coronal loop depends on a number of factors, including dissipation of the wave energy, pressure and temperature structure in the loop atmosphere, work action between the wave and a flow, and the sensitivity properties of the observing instrument.  For equilibrium loops, dissipation and pressure and temperature stratification dominate in the low corona where the waves are most easily detected.  Unfortunately, the effects are of comparable magnitude, and it would be difficult to disentangle them to infer the structure of an unknown loop atmosphere.  Coronal seismology offers no big advantage over spectroscopy in this regard.  

We have also shown that the intensity perturbations observed in 1-2 MK loops by {\it TRACE} are easily reproduced by propagating wave simulations with classical thermal conduction damping.  There is no need to invoke anomalously large conduction or a different damping mechanism altogether.  \cite{ow02} and \cite{mes04} similarly concluded that classical thermal conduction is adequate to explain the damping of longitudinal standing waves observed as Doppler shifts in much hotter ($> 6$ MK) loops.  The ``loops" in their simulations are non-radiating isothermal structures that are disconnected from the chromosphere and therefore do not include the complex energy balance of our simulations, but the basic conclusion concerning the effectiveness of classical thermal conduction damping should be correct.

Thermal conduction should not affect the transverse standing waves that are revealed as oscillating loops in {\it TRACE} movies \citep{sat02,aetal02}.  \cite{netal99} and \cite{oa02} suggested that these waves are instead damped by shear viscosity and/or electrical resistivity.  However, the magnitude of the damping must be at least $10^5$ times larger than predicted by classical transport theory.  Our study rules out the possibility that either thermal conduction or compressive viscosity is enhanced by a similar factor.  

It is not unreasonable that a mechanism of anomalous transport would affect the various transport processes differently.  This is known to be the case for classical transport whenever a strong magnetic field is present.  Cross-field shear viscosity is greatly inhibited compared to field-aligned compressive viscosity and thermal conduction.  The difference in the viscous coefficients is 10-12 orders of magnitude under coronal conditions.  Anomalous transport might also be highly anisotropic, though not necessarily in the same sense.  For example, turbulence would enhance shear viscosity much more than either compressive viscosity or thermal conduction (\citealt{b65}, p. 236).  We conclude that the damping of transverse waves by anomalous shear viscosity and the damping of longitudinal waves by classical thermal conduction are not inconsistent.  However, further detailed study of particle transport is called for, since there are many important applications throughout astrophysics and space physics.

We close by pointing out that the equilibrium models used in our study are not necessarily a good representation of the loops in which propagating waves are observed.  Those loops are very long and tend to occur at the perimeters of active regions.  A majority of more typical {\it TRACE} loops are known to be inconsistent with both static equilibrium \citep{asa01,wwm03} and steady flow equilibrium \citep{pkm03}, even if the heating is concentrated near the footpoints.  These loops are best explained as bundles unresolved strands that are heated impulsively and quasi-randomly by nanoflares \citep{c94,k02,wwm03b}.  In this picture, a given loop contains a wide distribution of strand temperatures at all times.  

Whether the long loops with waves are also multi stranded is not known. 
\cite{metal03} recently studied one of the cases in the \cite{dhiw2002a} sample and found that the intensity oscillations visible in {\it TRACE} 171 are simultaneously visible in lines of \ion{He}{1}, \ion{O}{5}, and \ion{Mg}{9} observed by the CDS instrument on {\it SOHO}.  This suggests that the loop may indeed be multi stranded.  If so, then waves launched at the base would propagate at different speeds in the different strands.  One consequence is that the waves would experience damping by phase mixing \citep{hp83,oa02}.  Another is that the wave front would become progressively distorted and appear to broaden with height, giving the impression of more damping than is actually present.  We plan to examine this latter possibility with additional modeling and a more detailed examination of the {\it TRACE} data.

\acknowledgments{We thank Alan Hood for suggesting this problem to us and Spiro Antiochos for helpful discussions.  We also thank the referee for useful comments.  This work was supported by NASA and the Office of Naval Research.}

\appendix
\section{Isothermal versus Adiabatic Waves}

\begin{figure}
\plotone{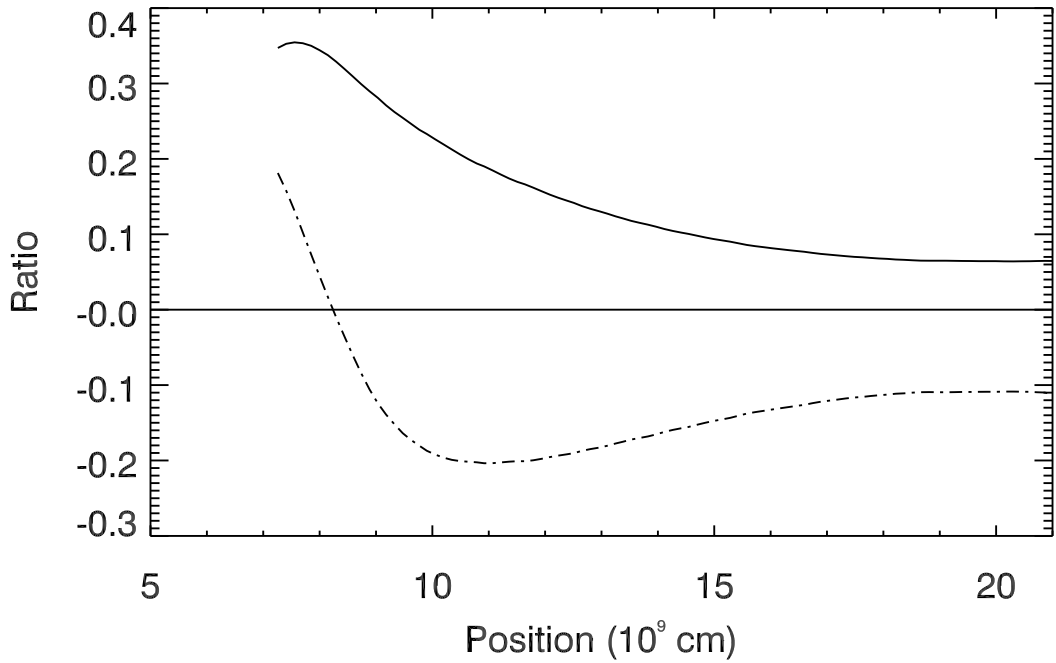}
\caption{Ratio of the left to right sides of Equation \ref{eq:13} (solid) and ratio of the second to first terms on the right-hand-side of Equation \ref{eq:11} (dot-dash) versus position along the loop in Case 1.  
\label{fig:isothermal}}
\end{figure}

Acoustic waves can behave isothermally, adiabatically, or somewhere in between depending on their wavelength and the temperature and density of the plasma in which they propagate.  For isothermal behavior, $\delta T = 0$, and for adiabatic behavior,
\begin{equation} 
\delta T = \left( \gamma - 1 \right) T \frac {\delta n}{n} . \label{eq:13}
\end{equation}
The solid line in Figure \ref{fig:isothermal} shows the ratio of the left to right sides of Equation \ref{eq:13} as a function of position along the loop for the waves simulated in Case 1, described in Section 4.  The ratio is $< 1$, but not $\ll 1$, indicating that the waves are roughly isothermal.    
This is consistent with the result of \citet[Eqn. C6a]{pks94a}, which can be modified to state that the transition between isothermal and adiabatic behavior occurs when 
\begin{equation}
\tau \approx \tau_{crit} = 65 \frac {T^{3/2}} {n} , \label{eq:tau}
\end{equation}
with shorter periods corresponding to isothermal behavior.  
The 250s period of the simulated waves is about 3 times shorter than $\tau_{crit}$ at $s = 10^{10}$ cm.  

The dashed line in Figure \ref{fig:isothermal} shows the ratio of the second to first terms on the right-hand-side of Equation \ref{eq:11}.  This ratio is of order 10 - 20 \% for most of the loop, justifying our neglect of the second term in arriving at Equation \ref{eq:12}.

\clearpage

\clearpage

\clearpage

\clearpage

\clearpage

\clearpage

\clearpage

\clearpage

\clearpage

\clearpage

\clearpage

\clearpage

\begin{deluxetable}{crrrrr}
\tablewidth{0pt}
\tablecaption{Observed Intensity Scale Lengths
\label{tbl-1}}
\tablehead{
\colhead{Loop} & \colhead{$\delta I_1$} & \colhead{$\delta I_2$} & 
\colhead{$\Delta s$} & \colhead{$H_{\delta I}^{obs}$} & 
\colhead{$\Delta H_{\delta I}^{obs}$}}
\startdata
2a &  6.73  & 8.19  & 1.02 &  -5.20 & 5.60  \\
3a &   15.90  &  7.99  & 0.58  & 0.88 & 0.25 \\
4a &  6.57  &  5.90  & 1.02 &  9.52 & 18.86 \\
4b &  7.72  &   8.42 & 0.44  &  -5.01 & 12.20 \\
4c &   4.81  &   4.69 & 1.16 &   48.39 & 427.61 \\
4d &   8.37 &  6.37 & 0.87 &   3.22  & 2.47 \\
5a &   14.17 &  7.93 & 0.44 &    0.77 & 0.27 \\
6a &    8.56 & 6.24  & 0.58 &    1.86 & 1.23 \\
7a &    7.05 & 6.44  & 0.29 &  3.18  & 7.36 \\
7b &   6.44 & 4.76  & 1.31 &  4.35  & 3.00 \\
7c &  6.61  &  4.40 & 0.44 &  1.09  & 0.55 \\
8a &   6.03  & 4.62  & 1.31 &   4.94 & 3.90 \\
9a &   10.15 & 6.08  & 0.73 &   1.45 & 0.57 \\
10a &  5.86 &  5.16 &  0.29 &   2.29  & 3.82 \\
11a &   11.63  &  8.46 &  0.73 &   2.30 & 1.50 \\
12a &  5.88  & 6.52 & 0.15 & -1.40 & 2.84 \\
13a &  2.85 &  3.28 & 1.02 & -7.25 & 10.93 \\
13b &  1.95 & 1.06 & 1.02 & 1.71 & 0.08\\
13c & 3.80 & 2.46 & 0.44 & 1.02 & 0.48 \\
13d &  2.72 & 2.88 & 0.44 & -7.98 & 30.98 \\
13e & 1.69 & 1.50 &  0.29 & 2.34 & 3.99 \\
13f &  1.57 & 1.80 & 1.16 & -8.77 & 13.99 \\
13g & 1.08 &  1.48 & 0.44 & -1.39 & 0.91 \\
14a &  3.22 & 1.84 & 0.73 & 1.33 & 0.50 \\
14b &  2.08 & 1.71 & 0.73 & 3.73 & 4.02 \\
14c & 5.96 &  3.81 & 1.74 & 3.96 & 1.82 \\
14d & 6.07 & 5.93 & 0.44 & 18.85 & 173.00 \\
14e &  7.82 &  12.17 & 0.73 & -1.67 & 0.78 \\
14f &  4.53 &  4.14 & 0.15 & 1.61 & 3.78 \\
15a &  3.79 &  4.93 & 0.58 & -2.23 & 1.79 \\
16a &   1.96 & 1.79  & 0.87 & 9.47 & 21.78 \\
16b &  2.34 &   1.21 & 0.58 & 0.91 & 0.27 \\
17a &  2.69 &  3.20 & 2.18 & -12.58 & 15.29 \\
17b &  6.10 &   7.59 & 0.44 & -2.00 & 1.93 \\
17c &  6.50 & 5.12  & 0.73 & 3.06 & 2.70 \\
17d &   3.40 &  4.78 & 0.44 & -1.29 & 0.78 \\
17e & 1.74  &  1.69 & 0.44 &  14.48 & 102.12 \\
\enddata
\tablecomments{Intensities are in units of counts s$^{-1}$ pix$^{-1}$, and lengths are in units of $10^9$ cm.}
\end{deluxetable}


\begin{thebibliography}{}

\bibitem[Antiochos {\it et al.}(1999)]{amsk99}
Antiochos, S. K., MacNeice, P. J., Spicer, D. S., \& Klimchuk, J. A. 1999, 
\apj, 512 985

\bibitem[Aschwanden {\it et al.} (2002)]{aetal02}
Aschwanden, M. J., De Pontieu, B., Schrijver, C, J., \& Title, A. M., \apj,206, 99

\bibitem[Aschwanden, Nightingale, \& Alexander (2001)]{ana00}
Aschwanden, M. J., Nightingale, R. W., \& Alexander, D. 2000, \apj,541, 1059

\bibitem[Aschwanden, Schrijver, \& Alexander (2001)]{asa01}
Aschwanden, M. J., Schrijver, C, J., \& Alexander, D. 2001, \apj,550, 1036

\bibitem[Berghmans \& Clette(1999)]{bc99}
Berghmans, D. \& Clette, F. 1999, \solphys, 186, 207


\bibitem[Braginskii (1965)]{b65} Braginskii, S. I. 1965, Rev. Plasma Phys., 1, 205

\bibitem[Cargill (1994)]{c94} Cargill, P. J. 1994, \apj, 422, 381

\bibitem[Cargill \& Klimchuk (1997)]{ck97} Cargill, P. J., \& Klimchuk, J. A. 1997, \apj, 478, 799

\bibitem[Dahlburg, Klimchuk, \& Antiochos (2004)]{dka04} Dahlburg, R. B., Klimchuk, J. A., \& Antiochos, S. K. 2004, \apj, submitted

\bibitem[Deforest \& Gurman(1998)]{dg1998}
Deforest, C. E. \& Gurman, J. B. 1998, \apjl, 501 L217

\bibitem[De Moortel \& Hood(2003)]{dh03}
De Moortel, I., \& Hood, A. W. 2003, \aap, 408, 755

\bibitem[De Moortel \& Hood(2004)]{dh04}
De Moortel, I., \& Hood, A. W. 2004, \aap, 415, 705

\bibitem[De Moortel {\it et al.}(2002)]{dhiw2002a}
De Moortel, I., Hood, A. W., Ireland, J., \& Walsh, R. W. 2002a, 
\solphys, 209 61


\bibitem[De Moortel, Ireland, \& Walsh(2000)]{diw2000}
De Moortel, I., Ireland, J., \& Walsh, R. W. 2000, \aap, 355 L23

\bibitem[De Moortel, Parnell, \& Hood(2003)]{dph03}
De Moortel, I., Parnell, C. E., \& Hood, A. W. 2003, \solphys, 215 69

\bibitem[Erdelyi, Ballester, \& Fleck(2004)]{ebf04}
Erdelyi, R., Ballester, J. L., \& Fleck, B. (ed.) 2003, ESA SP-547, Proceedings of the SOHO-13 Workshop on Waves, Oscillations, and Small-Scale Events in the Solar Atmosphere (Noordwijk: ESA)

\bibitem[Erdelyi {\it et al.}(2003)]{epra03}
Erdelyi, R., Petrovay, K., Roberts, B., \& Aschwanden, M. (ed.) 2003, NATO Science Ser. 124, Turbulence, Waves and Instabilities in the Solar Plasma (Dordrecht: Kluwer)

\bibitem[Heyvaerts \& Priest (1983)]{hp83}
Heyvaerts, J., \& Priest, E. R.  1983, \aap, 117, 220

\bibitem[Jacques (1977)]{j77}
Jacques, S. A. 1977, \apj, 215, 942

\bibitem[Klimchuk(2000)]{k2000}
Klimchuk, J. A. 2000, \solphys, 193 53

\bibitem[Klimchuk(2002)]{k02}
Klimchuk, J. A.  2002, Inst. for Theoretical Physics Conference on Solar Magnetism and Related Astrophysics, ed. G. Fisher \& D. Longcope (Santa Barbara: Univ. California) 
(http://online.kitp.ucsb.edu/online/solar\_c02/klimchuk/)

\bibitem[Klimchuk, Antiochos, \& Norton(2000)]{kan2000} 
Klimchuk, J. A., Antiochos, S. K., \& Norton, D. 2000, \apj, 542, 504

\bibitem[Klimchuk \& Cargill(2001)]{kc2001}
Klimchuk, J. A. \& Cargill, P. J. 2001, \apj, 553, 440

\bibitem[Klimchuk \& Gary(1995)]{kg95}
Klimchuk, J. A. \& Gary, D. E.  1995, \apj, 448, 925

\bibitem[Klimchuk {\it et al.}(1992)]{kea1992}
Klimchuk, J. A.,  Lemen, J. R., Feldman, U., Tsuneta, S., \& 
 Uchida, Y. 1992, PASJ, 44, L181 


\bibitem[Lenz {\it et al.} (1999)]{letal99}
Lenz, D. D., DeLuca, E. E., Golub, L., Rosner, R., \& Bookbinder, J. A. 1999, \apjl, 517, L155

\bibitem[Lopez-Fuentes \& Klimchuk(2004)]{lf2004}
Lopez-Fuentes, M., \$ Klimchuk, J. A. 2004, \apj, in preparation


\bibitem[Marsh {\it et al.}(2003)]{metal03}
Marsh, M. S., Walsh, R. W., De Moortel, I., \& Ireland, J. 2003, \aap, 404, L37

\bibitem[Mendoza-Briceno, Erdelyi, \& Sigalotti(2004)]{mes04}
Mendoza-Briceno, C. A., Erdelyi, R., \& Sigalotti, L. G. 2004, \apj, 605, 493

\bibitem[Meyer(1985)]{m85} Meyer, J.-P.  1985, \apjs, 57, 173

\bibitem[Nakariakov {\it et al.}(1999)]{netal99} 
Nakariakov, V. M., Ofman, L., DeLuca, E. E., Roberts, B., \& Davila, J. M. 1999, Science, 285, 862

\bibitem[Nakariakov {\it et al.}(2000)]{nvbr2000} 
Nakariakov, V. M.,  Verwichte, E., Berghmans, D., \& Robbrecht, E. 2000, \aap, 362 1151

\bibitem[Nightingale, Aschwanden, \& Hurlburt(1999)]{nah99}
Nightingale, R. W., Aschwanden, M. J., \& Hurlburt, N. E. 1999, \solphys, 190, 249

\bibitem[Ofman \& Aschwanden(2002)]{oa02}
Ofman, L., \& Aschwanden, M. J. 2002, \apjl, 576, L153

\bibitem[Ofman, Nakariakov, \& Sehgal(2000)]{ons00}
Ofman, L., Nakariakov, V. M., \& Sehgal, N. 2000, \apj, 533, 1071

\bibitem[Ofman {\it et al.}(1997)]{oetal97}
Ofman, L., Romoli, M., Poletto, G., Noci, C., \& Kohl, J. L. 1997, \apjl, 491, L111

\bibitem[Ofman, Nakariakov \& Deforest(1999)]{ond99}
Ofman, L., Nakariakov, V. M., \& Deforest, C. E. 1999, \apjl, 514, 441

\bibitem[Ofman \& Wang(2002)]{ow02}
Ofman, L. \& Wang, T. 2002, \apj, 580, L85

\bibitem[Patsourakos, Klimchuk \& MacNeice(2003)]{pkm03}
Patsourakos, S., Klimchuk, J. A., \& MacNeice, P. J.  2003, \apj, 603, 322

\bibitem[Porter, Klimchuk \& Sturrock(1994)]{pks94a}
Porter, L. J., Klimchuk, J. A., \& Sturrock, P. A. 1994, \apj, 435 482


\bibitem[Reale {\it et al.}(2000)]{rpsdg2000}
Reale, F.,  Peres, G., Serio, S., DeLuca, E. E., \& Golub, L. 2000, 
\apj, 535 412

\bibitem[Robbrecht {\it et al.}(2001)]{rvbhpn2001}
Robbrecht, E., Verwichte, E., Berghmans, D.,  Hochedez, J. F., 
Poedts, S.,  \& Nakariakov, V. M. 2001, \aap, 370, 591

\bibitem[Rosner, Tucker, \& Vaiana(1978)]{rtv78} Rosner, R., Tucker, W. H., 
\& Vaiana, G. S.  1978, \apj, 220, 643

\bibitem[Schrijver, Aschwanden, \& Title(2002)]{sat02}
Schrijver, C. J., Aschwanden, M. J., \& Title, A. M. 2002, \solphys, 206, 69

 

\bibitem[Van Leer(1979)]{v79}
Van Leer, B. 1979, J. Comput. Phys., 32, 101 

\bibitem[Wang {\it et al.} (2002)]{wetal02} 
Wang, T. J., Solanki, S. K., Curdt, W., Innes, D. E., \& Dammasch, I. E. 2002, \apjl, 574, L101

\bibitem[Warren, Winebarger, \& Mariska (2003)]{wwm03b} 
Warren, H. P., Winebarger, A. R.,  \& Mariska, J. T. 2003, \apj, 593, 1174

\bibitem[Watko \& Klimchuk(2000)]{wk2000}
Watko, J. A. \& Klimchuk J. A. 2000, \solphys, 193 77

\bibitem[Winebarger, Warren, \& Mariska (2003)]{wwm03} 
Winebarger, A. R., Warren, H. P., \& Mariska, J. T. 2003, \apj, 587, 439


\end{thebibliography}
\end{document}